# Ultra-low power nonlinear optics in a high Q CMOS compatible integrated micro-ring resonator

M. Ferrera,[1*] D. Duchesne,[1] L. Razzari,[1,2] M. Peccianti,[1] R. Morandotti,[1] P. Cheben,[3] S. Janz,[3] D.-X. Xu,[3] B. E. Little,[4] S. Chu,[4] and D. J. Moss[1,5**]

[1]*INRS-EMT, 1650 Boulevard Lionel Boulet, Varennes, Quebec, J3X-1S2, Canada*
[2]*Dipartimento di Elettronica, Universita` di Pavia, via Ferrata 1, 27100 Pavia, Italy*
[3]*Institute for Microstructural Sciences, National Research Council of Canada*
*1200 Montreal Road, Ottawa, Ontario K1A 0R6, Canada*
[4]*Infinera Corp, 9020 Junction Drive, Annapolis, Maryland 94089, USA*
[5]*CUDOS and IPOS, School of Physics, University of Sydney, New South Wales 2006, Australia*
*ferreram@emt.inrs.ca , **dmoss@physics.usyd.edu.au*

**Abstract:** We demonstrate efficient, low power, continuous-wave four-wave mixing in the C-band, using a high index doped silica glass micro ring resonator having a Q-factor of 1.2 million. A record high conversion efficiency for this kind of device is achieved over a bandwidth of 20nm. We show theoretically that the characteristic low dispersion enables phase-matching over a bandwidth > 160nm.

**1. Introduction**

Nonlinear optical frequency conversion is important for many applications, including all-optical switching, correlated photon pair generation, and narrow linewidth or multi-wavelength sources [1-5]. Four-wave mixing (FWM), a third order nonlinear parametric process, is a common approach to frequency conversion, and optimizing its efficiency is critical to realizing practical levels of performance. This can be achieved either by increasing the waveguide nonlinearity parameter $\gamma = \omega n_2/c \cdot A_{eff}$ (where $A_{eff}$ is the effective area of the waveguide, $c$ is the speed of light, $n_2$ is the Kerr nonlinearity, and $\omega$ is the pump angular frequency), or by using resonant structures to enhance the local field intensity. Extremely high $\gamma$'s have been reported in silicon [6] (200,000 $W^{-1} km^{-1}$) and chalcogenide glass [7] (100,000 $W^{-1} km^{-1}$) nanowaveguides. However, a low intrinsic figure of merit ($n_2/(\alpha_2 \cdot \lambda)$, where $\alpha_2$ is the two-photon absorption coefficient) in silicon [8,9] as well as an immature fabrication process and possible material power handling issues in chalcogenides, have motivated the search for additional platforms. High Q cavities have been effective in enhancing FWM in silicon [10] and GaAs [11, 12] integrated devices, as well as in silica [5, 13] and chalcogenide glass [14-16] microspheres and toroids.

Last year [17], we reported the first demonstration of continuous-wave (CW) nonlinear optics in high index doped silica glass waveguides, by achieving wavelength conversion via FWM in an integrated ring resonator (Q=65,000) with only a few mW of pump power. This established the capability of robust high index glasses, with negligible nonlinear absorption [18-21], to provide a platform for all-optical nonlinear photonic integrated circuits, with the added benefit of CMOS compatible (low temperature) fabrication processes. More recently [22], similar characteristics and performances have also been reported in silicon nitride, which closely resembles high index doped silica glass in terms of both nonlinearity and index.

Here, we report CW FWM in an integrated ring resonator with Q factor of 1.2 million, the highest Q-factor to date for a monolithically integrated ring resonator in which nonlinear optics has been reported. This high Q resonator is ideal for applications other than telecommunications, such as narrow linewidth, multi-wavelength sources, or correlated photon pair generation [1]. We achieve an improvement in efficiency of 23dB over previous results [17], with negligible phase mismatch for a conversion bandwidth (signal to idler separation) of 20nm. In addition, we experimentally measured an extremely low dispersion in these waveguides and found that it is nearly ideal for FWM in the C-band. Using these results we predict an achievable tuning range of 160nm. The combination of high efficiency, wide tuning range, and very low linear and nonlinear losses in these structures together with well established fabrication processes that do not require high temperature annealing (CMOS compatible), indicate that this platform is highly promising for a wide range of applications.

The four wave mixing is a $\chi^{(3)}$ (third order susceptibility) parametric process where two pump photons mix with a signal photon to yield an idler photon, whose frequency is established through energy conservation:

$$2\upsilon_p = \upsilon_s + \upsilon_i, \qquad (1)$$

where $\upsilon_p$, $\upsilon_s$ and $\upsilon_i$ are the pump, signal, and idler frequencies respectively. This process can also occur spontaneously, without seeding by a separate input at frequency $\upsilon_s$, allowing the possibility of correlated photon pair generation. The classical effect (with seeding) is stronger and is the basis of numerous all-optical signal processing schemes in integrated devices and optical fibers [6, 23, 24]. To take advantage of the resonant enhancement for FWM in a ring resonator, the pump and signal frequencies are typically aligned to resonances, and if the process is phase matched, the idler wave will also be resonant, yielding a tremendous enhancement in efficiency.

## 2. Experiment

The high Q ring resonator we studied has a radius $R=135\mu m$ and a free spectral range (FSR) of 200GHz (see Fig. 1a). The ring is vertically coupled to bus waveguides, both of which are composed of high-index $n=1.7$ doped silica glass (Hydex®) [17]. The linear transmission spectrum was measured from 1460nm to 1630nm with an Agilent 81600B CW tunable laser (with an 81619A photodetector). Spectra were taken with a resolution of 0.3pm, a balance between managing thermal stability and accurately measuring the 1.3pm (or 160MHz) linewidth of the device (Fig. 1b). This linewidth is equivalent to a Q-factor of 1.2 million, the highest reported for a *monolithically* integrated device (ie., excluding micro-toroids and spheres). Since the resonance can shift by as much as a half-width with a temperature change of only ~ 0.1C, the chip was mounted on a temperature controlled stage with thermocouple feedback in order to maintain alignment of the resonances to the optical wavelengths.

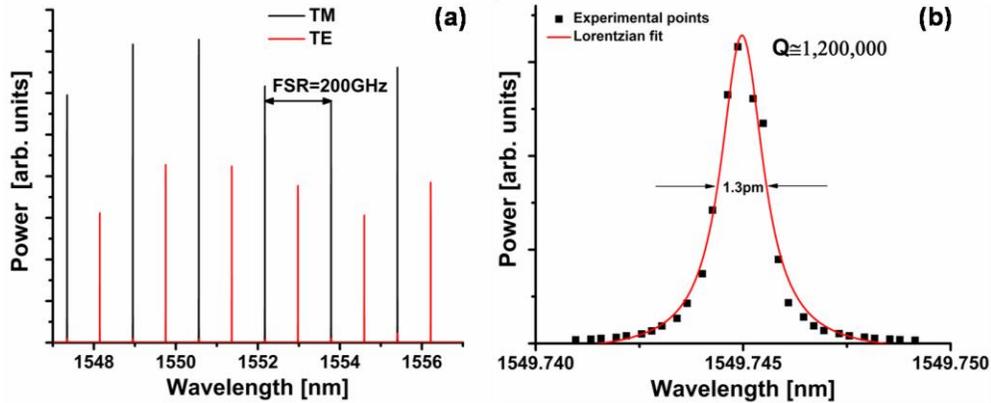

Fig. 1. Linear transmission spectrum of the ring resonator recorded at the Drop port (a) and high resolution detail for a typical resonance (b) (with a lorentzian fit) showing a resonance width of 1.3pm. The variation in resonance strengths in a) are mainly a result of the 0.3pm spectral resolution.

The dispersion of the ring resonator can be determined from the experimentally measured resonance wavelengths using

$$m\lambda_m = n_m L = n_m 2\pi R, \tag{2}$$

where $\lambda_m$ is the wavelength of the $m^{th}$ resonance with effective index $n_m$. Using modal simulations ($n$=1.7 for the waveguide core, $n$=1.444 for $SiO_2$ @ $\lambda$=1550nm) we extract the mode number $m$ of a specific resonance and from Eq. (2) the mode propagation constant $\beta$ can then be obtained at each resonance $m$:

$$\beta_m \equiv \frac{n_m \omega_m}{c} = \frac{m}{R} \tag{3}$$

using the experimentally measured resonant frequencies from 1460-1630nm. To obtain the dispersion we fit the propagation constant to a 4$^{th}$ order polynomial and took the second order derivative with respect to the angular frequency.

For the FWM experiments (Figure 2) the pump and signal beams from two CW tunable lasers (Agilent model 8164A and 8164B) were tuned to resonances of the ring resonator, then passed through two fiber polarization controllers and then coupled into the "on-chip" ring resonator via fiber pigtails. The output of both the Drop and Through ports were then measured via an optical spectrum analyzer (OSA) and power meter respectively.

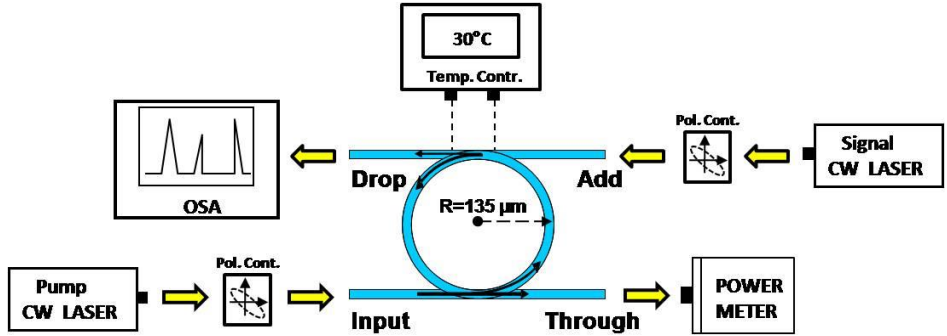

Fig. 2. FWM experimental setup.

## 3. Results and Discussion

The resultant group velocity dispersion is shown in Fig. 3 for both the quasi- TE and TM modes. The dispersion for both modes is small and anomalous over most of the C-band. At 1550nm we obtain an anomalous GVD of $\beta_2$= -3.1 ± 0.9 ps$^2$/km ($D$ = 2.5 ± 0.7 ps/nm/km) for the TM mode and -10 ± 0.9 ps$^2$/km ($D$ = 8 ± 0.7 ps/nm/km) for the TE mode. The zero dispersion points were determined to be at 1560nm and 1595nm for the TM and TE modes respectively. These results are promising for FWM in the telecommunications bands, as a low dispersion implies low phase mismatch, and hence large conversion efficiency.

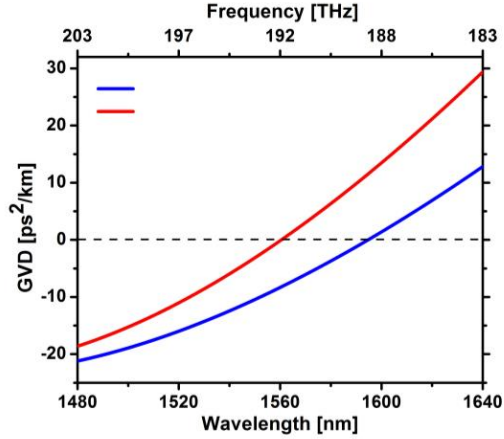

Fig. 3. Group velocity dispersion of the resonator obtained by fitting the experimentally measured resonance frequencies. The fits include up to 4th order dispersion terms.

The results for two FWM experiments are presented in Fig. 4, which shows the spectrum out of the Drop port. In the first experiment, the pump and signal wavelengths were offset by one FSR (200GHz) near 1550nm, while in the second experiment the pump was shifted to ~ 1600nm and the signal was six FSRs away (1.2THz). With incident pump and signal powers of 8.8mW and 1.25mW, respectively, we obtained an external conversion efficiency (defined as the ratio of the output idler power to the input signal power, thus accounting for all coupling, insertion and propagation losses) of ~ -36dB for both experiments.

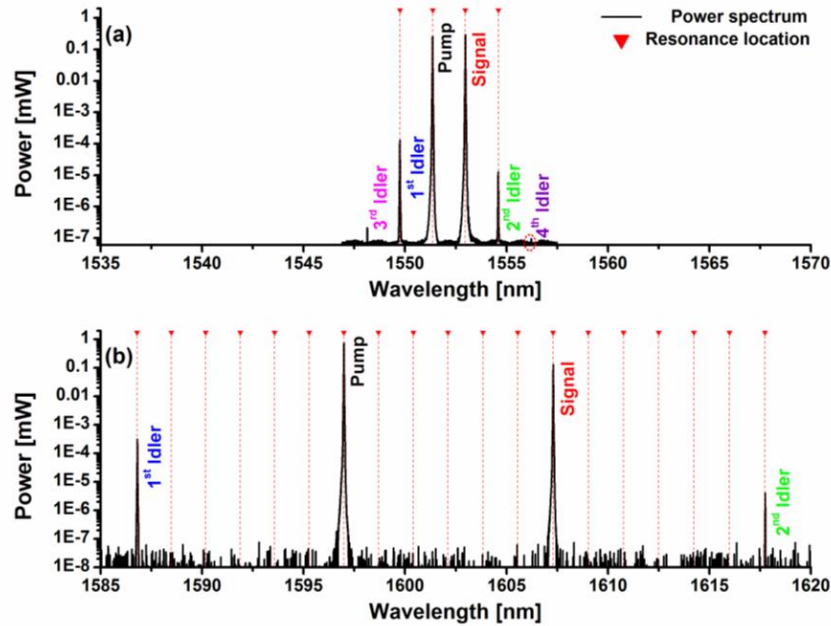

Fig. 4. FWM experimental results for pump and signal wavelengths tuned to adjacent esonances for the TE mode (200GHz) (a) and 6 resonances apart (~1.2THz) for the TM mode (b). The 3rd and 4th idlers (a) demonstrate the onset of cascaded FWM. Note the linewidths are broader than in Figure 1 because of the limited OSA resolution (15pm).

The internal efficiency is estimated to be -26dB for both experiments using coupling losses from [17, 18], which is 23dB higher than previously reported in ring resonators [17]. This agrees with the theoretical undepleted pump model which predicts [12]:

$$\eta = \frac{P_{Idler}}{P_{Signal}} = |L \cdot \gamma|^2 \cdot (FE)^8 \cdot P_{Pump}^2, \tag{6}$$

where $\eta$ is the internal conversion efficiency, $L$ is the resonator length, the $P$'s are the optical powers, and $FE$ is the field enhancement factor [17] estimated from the resonator geometry. Our high conversion efficiency is thus seen to be a direct result of the field enhancement factor in this high Q-factor resonator ($FE$=17.9). The similar efficiencies we obtained in the two FWM experiments, despite the different pump wavelengths and pump/signal wavelength offsets, is a reflection of the very low net dispersion, which is also responsible for the onset of the cascaded FWM observable in Fig. 4a.

The phase matching condition for FWM in ring resonators can expressed in terms of the requirement matching the idler frequency ( Eq. (1)) to a ring resonance; i.e.:

$$\Delta \upsilon = |\upsilon_i - \upsilon_{res}| < \frac{\Delta \upsilon_{FWHM}}{2}, \tag{7}$$

where $\Delta\upsilon_{FWHM}$ is the resonance linewidth. Figure 5 shows surface plots of this frequency detuning as a function of pump frequency and pump / signal frequency offset, obtained from the experimentally measured dispersion. The black diagonal band in Fig. 5 reflects the range over which our experimental data were taken; the colored regions are where FWM is phase matched according to Eq. (7), whereas the black region is where it is not phasematched. In particular, for a pump frequency near 193THz, we expect a tuning range of > ±10THz, or ±80nm (160nm full range) for the quasi-TE polarized mode. This tuning range is remarkable given the extremely narrow linewidth of our ring resonators (160MHz), and is a direct result of the zero-GVD crossings.

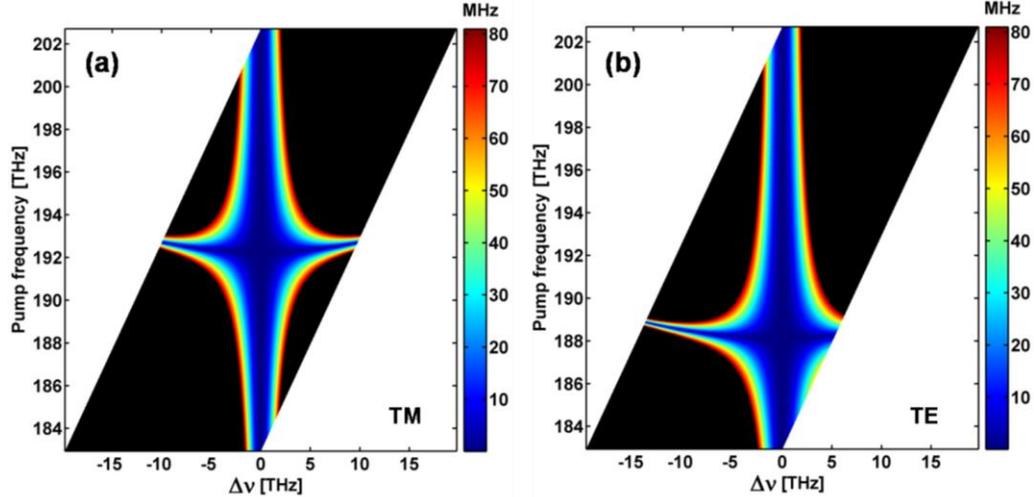

Fig. 5. Idler detuning (see Eq. (7)) for TM (a) and TE (b) polarizations. Δν is the difference between pump and signal frequencies (both independently tuned to resonances). Black regions represent non-phase matched conditions (where the idler is further than a FWHM from a central resonance frequency), whereas colored regions represent varying degrees of phase matching, with blue representing ideal phase matching. The black band reflects the region where experimental data was taken for the dispersion measurements.

## 4. Conclusions

We have demonstrated efficient FWM in an integrated, high index doped silica glass ring resonator with a Q-factor of 1.2 million. We obtain an external conversion efficiency 23dB higher than previous results, over a bandwidth of ±10nm. The waveguide dispersion is measured and found to be small and anomalous over most of the C-band, and we use this to predict a tuning range of ±80nm for a pump near 1550nm. Together with the reliability and CMOS compatible fabrication process of these waveguides, our results open up the possibility of efficiently performing all-optical frequency conversion for a wide range of applications such as narrow linewidth, and/or multi-wavelength sources, as well as the on-chip generation of correlated photon pairs.

## 5. Acknowledgements

This work was supported by the Australian Research Council Centres of Excellence program, the FQRNT (Le Fonds Québécois de la Recherche sur la Nature et les Technologies), the Natural Sciences and Engineering Research Council of Canada (NSERC), NSERC Strategic and Discover Projects and the INRS. L. R. and M. P. wishes to acknowledge a Marie Curie Outgoing International Fellowship contract no. 040514 and TOBIAS-PIOF-GA-2008-221262 respectively. We are also thankful to Edith Post for technical assistance.

**References and links**

1. P. Kolchin, S. Du, C. Belthangady, G.Y. Yin, S. E. Harris, and E. L. Ginzton "Generation of narrow-bandwidth paired photons: use of a single driving laser," Phys. Rev. Lett. **97**, 113602 (2006).
2. D. Klonidis, C. T. Politi, R. Nejabati, M. J. O'Mahony, and D. Simeonidou, "OPSnet: design and demonstration of an asynchronous high-speed optical packet switch," J. Lightw. Tech. **23**, 2914-2925 (2005).
3. J. Ma and C. Jiang, "Design and analysis of all-optical switches based on fiber parametric devices," Opt. Comm. **281**, 2605–2613 (2008).
4. K. Kawase, J. Shikata, K. Imai, and H. Ito, "Transform-limited, narrow-linewidth, terahertz-wave parametric generator," Appl. Phys. Lett. **78**, 2819-2821 (2001).
5. K. J. Vahala, "Optical microcavities," Nature **424**, 839-846 (2003).
6. R. Salem, M. A. Foster, A. C. Turner, D. F. Geraghty, M. Lipson, and A. L. Gaeta, "Signal regeneration using low-power four-wave mixing on silicon chip," Nature Photon. **2**, 35-38 (2008).
7. E. C. Mägi, L. B. Fu, H. C. Nguyen, M. R. E. Lamont, D. I. Yeom, and B. J. Eggleton, "Enhanced Kerr nonlinearity in sub-wavelength diameter $As_2Se_3$ chalcogenide fiber tapers," Opt. Expr. **15**, 10324-10329 (2007).
8. L. Yin, and G. P. Agrawal, "Impact of two-photon absorption on self-phase modulation in silicon Waveguides," Opt. Lett. **32**, 2031–2033 (2007).
9. G. Priem, P. Bienstman, G. Morthier, and R. Baets, "Impact of absorption mechanisms on Kerr-nonlinear resonator behavior," J. Appl. Phys. **99**, 063103 (2006).
10. A. C. Turner, M. A. Foster, A. L. Gaeta, and M. Lipson, "Ultra-low power parametric frequency conversion in a silicon microring resonator", Opt. Express **16**, 4881-4887 (2008).
11. J. E. Heebner, N. N. Lepeshkin, A. Schweinsberg, G. W. Wicks, R. W. Boyd, R. Grover, and P. T. Ho, "Enhanced linear and nonlinear optical phase response of AlGaAs microring resonators," Opt. Lett. **29**, 769-771 (2004).
12. P. P. Absil, J. V. Hryniewicz, B. E. Little, P. S. Cho, R. A. Wilson, L. G. Joneckis, and P.-T. Ho, "Wavelength conversion in GaAs micro-ring resonators," Opt. Lett. **25**, 554-556 (2000).
13. I. H. Agha, Y. Okawachi, M. A. Foster, J. E. Sharping, and A. L. Gaeta, "Four-wave-mixing parametric oscillations in dispersion-compensated high-*Q* silica microspheres," Phys. Rev. A **76**, 043837 (2007).
14. D. H. Broaddus, M. A. Foster, I. H. Agha, J. T. Robinson, M. Lipson, and A. L. Gaeta, "Silicon-waveguide-coupled high-Q chalcogenide microspheres," Opt. Express **17**, 5998-6003 (2009).
15. D Freeman, C Grillet, MW Lee, CLC Smith, Y Ruan, A Rode, M.Krolikowska, S.Tomljenovic-Hanic, C Martijn De Sterke, MJ Steel, B Luther-Davies, S Madden, D J Moss, Y Lee, and B J Eggleton, "Chalcogenide glass photonic crystals", Photonics and Nanostructures-Fundamentals and Applications **6**, (1) 3-11 (2008).
16. C Grillet, C Monat, C Smith, B J Eggleton, DJ Moss, S Frédérick, D Dalacu, P Poole, J Lapointe, G Aers, and RL Williams, "Nanowire coupling to photonic crystal nanocavities for single photon sources", Optics Express **15**, (3) 1267-1276 (2007).
17. M. Ferrera, L. Razzari, D. Duchesne, R. Morandotti, Z. Yang, M. Liscidini, J. E. Sipe, S. Chu, B. E. Little, and D. J. Moss, "Low-power continuous-wave nonlinear optics in doped silica glass integrated waveguide structures," Nature Photonics **2**, 737-740 (2008).
18. D. Duchesne, M. Ferrera, L. Razzari, R. Morandotti, S. Chu, B. Little, and D. J. Moss, "Efficient self-phase modulation in low loss, high index doped silica glass integrated waveguides," Optics Express **17**, 1865-1870 (2009).

19. A Pasquazi, Y Park, J Azaña, F Légaré, R Morandotti, BE Little, ST Chu, and DJ Moss, "Efficient wavelength conversion and net parametric gain via four wave mixing in a high index doped silica waveguide", Optics Express **18**, (8), 7634-7641 (2010).
20. D Duchesne, M Peccianti, MRE Lamont, M Ferrera, L Razzari, F Légaré, R Morandotti, S Chu, BE Little, DJ Moss, "Supercontinuum generation in a high index doped silica glass spiral waveguide", Optics Express **18,** (2) 923-930 (2010).
21. M Peccianti, M Ferrera, L Razzari, R Morandotti, BE Little, ST Chu, and DJ Moss, "Sub-picosecond optical pulse compression via an integrated nonlinear chirper", Optics Express **18,** (8) 7625-7633 (2010).
22. J. S. Levy, A. Gondarenko, M. A. Foster, A. C. Turner-Foster, A. L. Gaeta, and M. Lipson, "CMOS-compatible multiple-wavelength oscillator for on-chip optical interconnects", Nature Photonics **4,** (1) 37-40 (2010).
23. B Corcoran, TD Vo, MD Pelusi, C Monat, DX Xu, A Densmore, R Ma, S Janz, DJ Moss, and BJ Eggleton, "Silicon nanowire based radio-frequency spectrum analyzer", Optics Express **18,** (19) 20190-20200 (2010).
24. M. D. Pelusi, F. Luan, E. Magi, M. R. E. Lamont, D. J. Moss, B. J. Eggleton, J. S. Sanghera, L. B. Shaw, and I. D. Agrawal, "High bit rate all-optical signal processing in a fiber photonic wire," Opt. Expr. **16**, 11506-11512 (2008).